\documentclass[aps,prl,reprint,groupedaddress]{revtex4-1}
\usepackage{graphicx,subfigure}
\usepackage{amsmath,latexsym}

\newcommand{\ket}[1]{\left| #1 \right\rangle}
\newcommand{\bra}[1]{\left\langle #1 \right|}
\def\inn#1#2{\left\langle{#1}|{#2}\right\rangle}

\begin{document}

\title{Onset of the Limit Cycle and Universal Three-Body Parameter in Efimov Physics}
\author{Yusuke Horinouchi}
\author{Masahito Ueda}
\affiliation{Department of Physics, University of Tokyo, Hongo 7-3-1, Bunkyo-ku, Tokyo 113-8654, Japan}

\date{\today}

\begin{abstract}
The Efimov effect is the only experimentally realized universal phenomenon that exhibits the renormalization-group limit cycle with the three-body parameter parametrizing a family of universality classes. Recent experiments in ultracold atoms have unexpectedly revealed that the three-body parameter itself is universal when measured in units of an effective range. By performing an exact functional renormalization-group analysis with various finite-range interaction potentials, we demonstrate that the onset of the renormalization-group flow into the limit cycle is universal, regardless of short-range details, which connects the missing link between the two universalities of the Efimov physics. A close connection between the topological property of the limit cycle and few-body physics is also delineated.
%We have performed an exact functional renormalization-group (FRG) analysis of the Efimov physics with various finite-range interaction potentials. At high energy, the RG flow of the three-body coupling constant $g_3$ exhibits different behavior which depends on the short-range details of each individual interaction potential; however, in the infrared regime, it approaches to the limit cycle, the onset of which is found to give the universal value of the three-body parameter $a_-/r_{\mathrm{eff}}=-4.3$ in units of the effective range $r_{\mathrm{eff}}$.
%The universal three-body parameter in the Efimov effect is among the greatest recent experimental discoveries in the field of ultracold atoms. This experimental finding has been numerically vindicated [J. Wang {\it et al}., Phys. Rev. Lett. {\bf 108}, 263001 (2012)] and its physical origin has been clarified [P. Naidon {\it et al}., Phys. Rev. Lett. {\bf 112}, 105301 (2014)]. A major missing link in this subject is a relation between the universal three-body parameter and the fundamental property of the Efimov effect, i.e. a renormalization-group limit cycle. In this Letter, we show that they are in fact closely connected, by means of the functional renormalization-group method, the non-perturbative character of which plays a decisive role in revealing this connection.
\end{abstract}

% insert suggested PACS numbers in braces on next line
% insert suggested keywords - APS authors don't need to do this
%\keywords{}

%\maketitle must follow title, authors, abstract, \pacs, and \keywords
\maketitle

In the early 1970s, V. Efimov predicted a counterintuitive quantum phenomenon, in which resonantly interacting three bosons form an infinite series of three-body bound states even if the interaction is too weak to support a two-body bound state \cite{PLB-33-563}. The Efimov effect emerges in a wide range of systems including identical bosons \cite{PLB-33-563}, mass-imbalanced fermions \cite{NPA-210-157,PRA-67-010703}, particles in mixed dimension \cite{PRA-79-060701}, nucleons \cite{ARNPC-60-207}, magnons \cite{NP-9-93}, and macromolecules such as DNAs \cite{PRL-110-028105}.
Besides its universality, the Efimov spectrum shows a discrete scale invariance, where the energy eigenvalues of the trimers are related to one another by a universal scaling factor of $22.7^2$. This peculiar property provides a unique example of a renormalization-group (RG) limit cycle \cite{PRD-3-1818}, which refers to a periodic behavior of a RG flow and had been elusive until the emergence of the Efimov effect. Because of the universality and uniqueness, the Efimov effect has been extensively studied in various fields of physics such as atomic, chemical, nuclear, and particle physics. In particular, experimental observations of the Efimov effect in ultracold atoms \cite{N-440-315,NP-5-227,NP-5-586,PRL-103-163202,S-326-1683,PRL-102-165302} have given an enormous impetus to the development of Efimov physics. \par
\begin{figure}[tb]
  			\begin{center}
  			 	\includegraphics[width=200pt]{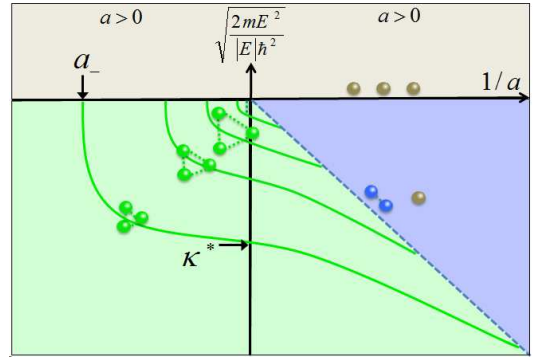}
				\caption{Energy spectrum of three-identical bosons with resonant interaction (not to scale). The abscissa shows the inverse $s$-wave scattering length $a^{-1}$ and the ordinate shows the square root of the energy eigenvalues. The Efimov states are related to one another by the universal scaling factor of $e^{2\pi/s_0}\simeq 22.7^2$, with $s_0\simeq1.00624$. The three-body parameters which set the energy scale of the lowest-lying Efimov state are labeled in this figure as $a_-$ and $\kappa^*$. Here $a_-$ denotes the $s$-wave scattering length at the triatomic resonance and $\kappa^*$ is the binding wave number of the lowest-lysing Efimov state at unitarity. They are uniquely related to each other.}\label{fig:esp}
  			\end{center}
\end{figure}

Among the crucial discoveries in recent ultracold atom experiments is the universality in the three-body parameter $\kappa^*$ (or equivalently the scattering length $a_-$ at the triatomic resonance) \cite{PRL-107-120401,PRL-108-145305,PRL-111-053202}, which sets the energy scale of the lowest-lying Efimov state (see Fig.~\ref{fig:esp}). While low-energy two-body observables are universally described by  the $s$-wave scattering length $a$, the existence of Efimov states leads to an additional dependence of the low-energy three-body observables on the three-body parameter, which encapsulates short-range details of the three-body physics and had therefore been considered to be non-universal.
%While low-energy two-body observables are universally described by a single parameter, $s$-wave scattering length $a$, the existence of the Efimov states leads to an additional dependence of the low-energy three-body observables on the three-body parameter $\kappa^*$ (or equivalently the triatomic resonance point $a_-$), which characterize the energy eigenvalue of the lowest Efimov state. Since $\kappa^*$ ($a_-$) encapsulate all details of interactions which are irrelevant at low energy, the effect of the Efimov states enter in low-energy three-body observables through this single parameter. However, because of its connection to short-range details of interactions, the precise value of $\kappa^*$ ($a_-$) had been regarded as non-universal and believed to change according to the details of interaction potential. \par
Recent experiments in ultracold atoms, however, have revealed that $a_-$ takes on almost the same value when measured in units of the van der Waals length $r_{\mathrm{vdW}}$ for various atomic species, different internal states, and different Feschbach resonances, suggesting some underlying physics that makes such an agreement possible. Recently, it has been suggested \cite{PRL-108-263001,PRA-86-052516,arXiv-1208-3912} that systems other than the atomic van der Waals systems such as nuclear systems fall into a similar universality class if the three-body parameter is measured in units of the effective range which characterizes the range of interactions.\par
%rg study of efimov: its importance
In this Letter, we address an as yet unexplored fundamental question: Does this experimentally found universality bring about a new universality in the renormalization-group limit cycle? We answer this question in the affirmative by focusing on identical bosons at the unitarity limit $a=\pm\infty$. Furthermore, how topological properties of a renormalization group flow can be related to universal aspects of few-body physics is an intriguing but so far untapped problem. We will touch upon this point at the end of this Letter, taking four-body physics as an illustrative example. \par
Previous RG analyses \cite{PRL-82-463,AP-321-306,PRA-79-042705} of the Efimov effect have been limited to the zero-range model $V_z(\mathbf{r})=g\delta(\mathbf{r})$, in which $\kappa^*$ is treated as an input parameter, precluding any statement about its universality. Since the finite-range effect plays an essential role in this universality, we should extend the RG analysis to systems with finite-range interactions. In previous works, the zero-range model was adopted to discuss the limit cycle; however, it predicts the periodic RG flow not only in the infrared limit but also in the ultraviolet limit, which indicates the unphysical Thomas collapse \cite{PR-47-903}.\par
\begin{figure}[tb]
  			\begin{center}
  			 	\includegraphics[width=250pt]{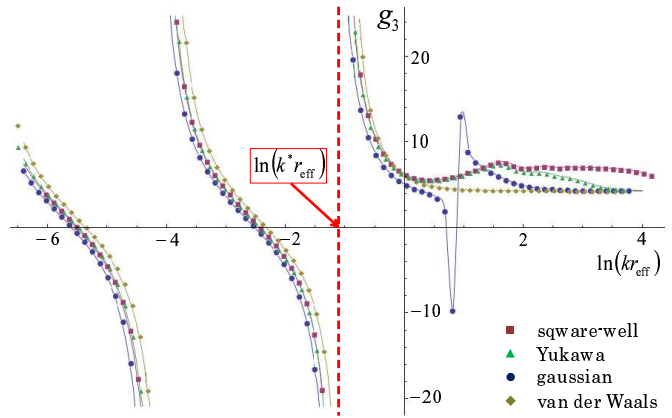}
				\caption{(color online) Exact RG flows of the three-body coupling constant $g_3$. The abscissa shows the logarithm of the RG scale $\mathrm{ln}(kr_{\mathrm{eff}})$ in units of the effective range $r_{\mathrm{eff}}$, and the ordinate shows the three-body coupling constant $g_3$. Different curves correspond to different interaction potentials: Gaussian ($\bigcirc$), van der Waals ($\Diamond$), Yukawa ($\triangle$), and sqware-well ($\Box$) potentials. The universal onset point of the limit cycle coincides with the universal value of the three-body parameter $\kappa^*r_{\mathrm{eff}}=0.49$.}\label{fig:result}
  			\end{center}
\end{figure}

In contrast, since we are interested in the universality of the three-body parameter, in which a finite-range nature of the interaction plays a crucial role \cite{PRL-108-263001,PRA-86-052516,arXiv-1208-3912,PRL-112-105301}, we have performed a functional renormalization-group (FRG) analysis for different Hamiltonians with various finite-range interactions. We have obtained an exact RG flow of the three-body coupling constant, which is defined as a dimensionless particle-dimer scattering amplitude (see Fig.~\ref{fig:result}). We have found that, in contrast with the zero-range model, the RG flow starts at a point away from the limit cycle and exhibits characteristic behavior that depends on the short-range details of each individual interaction potential; however, in the infrared regime, the flow begins to show the limit-cycle behavior, the onset of which is found to give the universal value of the three-body parameter $\kappa^*r_{\mathrm{eff}}=0.49$. The onset is evaluated as the RG scale at which the first divergence of the coupling constant occurs. By using the universal relation between $\kappa^*$ and $a_-$ \cite{PRA-86-052516}, we obtain $a_-=-4.3r_{\mathrm{eff}}$ in excellent agreement with the experimental results \cite{PRL-107-120401,PRL-108-145305,PRL-111-053202}. We thus identify the universality of the onset of the limit cycle with that of the three-body parameter. We note that the non-perturbative nature of the FRG have played a decisive role in revealing this relation, since we have to deal with a diverging coupling constant which the perturbative Wilsonian RG cannot deal with. It is striking that the geometrical property (i.e., the onset point) of the limit cycle can be related to the universality of the three-body parameter in the Efimov physics. This observation can be further generalized to the topological constraint of the limit-cycle behavior on the relationship between an Efimov state and its four-body companions as we discuss later.\par

%Since the discrete scale invariant Efimov states can be understood as a limit-cycle behavior of the renormalization-group flow, it is natural to consider that the energy scale at which the limit cycle starts characterizes the three-body parameter, thereby that the universal three-body parameter can be understood as a universal starting point of the limit cycle. \par
% Although recent theoretical works have already suggested that the finiteness of the range of interactions plays an essential role in this universality, the renormalization-group study of the Efimov physics is limited to the zero-range model, which treats the three-body parameter as a free parameter, preventing any statements about its universality. One reason is that it is quite hard to perform a renormalization-group analysis for realistic interactions with finite range, because of the non-perturbative nature of the Efimov effect and the complex momentum dependence of flowing coupling constants.
We now present our theoretical framework for obtaining these results. To perform an exact RG analysis on finite-range interactions, we use a simple microscopic model which accurately reproduces pair correlations of model interaction potentials and can be solved exactly for three particles. We use a separable-potential model whose interaction Hamiltonian is written in the form of a projection operator $\hat{V}_f=\xi\ket{\chi}\bra{\chi}$, which retains the simplicity of the zero-range (delta-function) interaction $\hat{V}_z=g\ket{\mathbf r}\bra{\mathbf r}$. The microscopic action for identical bosons is then written as
\begin{eqnarray}
	\nonumber S[\psi,\psi^*]=&&\int_{Q}\psi^*(Q)(i{q^0}+{q}^2-{\mu})\psi(Q) \\
	\nonumber&&+\frac{\xi}{4}\int_{Q_1Q_2Q_1'Q_2'}\delta({Q_1}+{Q_2}-{Q_1}'-{Q_2}')\\
	\nonumber&&\quad\times\chi\left(\frac{{\mathbf q}_1'-{\mathbf q}_2'}{2}\right)
		\chi^*\left(\frac{{\mathbf q}_2-{\mathbf q}_1}{2}\right)\\
	&&\quad\times\psi^*({Q_1}')\psi^*({Q_2}')\psi({Q_2})\psi({Q_1}),
\end{eqnarray}
where $Q$ denotes the four momentum consisting of Matsubara frequency $q_0$ and momentum ${\mathbf q}$, $\mu$ is the chemical potential, $\chi(q):=\inn{\mathbf{q}}{\chi}$ in momentum representation, $\psi$ denotes the bosonic field, and $\int_Q=\int\frac{{\mathrm d}q^0{\mathrm d}^3q}{(2\pi)^4}$. Throughout this Letter, we employ the units $\hbar=k_{\mathrm{B}}=2m=1$, where $k_{\mathrm{B}}$ is the Boltzmann constant and $m$ is the mass of the particle.\par
We can choose an appropriate $\ket{\chi}$ of $\hat{V}_f$ so that $\hat{V}_f$ reproduces the low-energy pair correlation, including a nonzero range, of model potentials. This approximation can be systematically developed with arbitrarily high accuracy by adding another projection term to $\hat{V}_f$ \cite{PRC-8-46}. The construction procedure of $\ket{\chi}$ is described in Refs.~\cite{PRC-8-46,PRL-112-105301}.
%The construction procedure of $\ket{\chi}$ is summarized as follows: First, for a given model potential, we calculate the zero-energy two-body $s$-wave radial wave function $\psi_0(r)$ with the asymptotic limit $1-r/a$, where $a$ is the scattering length. Then, we construct a separable potential which reproduces $\psi_0(r)$ exactly through following formula:
%\begin{eqnarray}
%&\displaystyle\chi(q)=1-q\int_0^{\infty}\dm r\left( 1-\frac{r}{a}-\psi_0(r)\right)\mathrm{sin}\left(qr\right),\\
%&\displaystyle\xi=4\pi\left(\frac{1}{a}-\frac{2}{\pi}\int_0^{\infty}\dm q|\chi(q)|^2\right)^{-1},
%\end{eqnarray}
%where $a$ is the $s$-wave scattering length.
Despite its simplicity, $\chi(q)$ reproduces two-body observables, including phase shifts and bound-state energies, of exact model potentials with high accuracy. Here we use four different types of the separable models: van der Waals, Yukawa, infinite square-well, and Gaussian, which are available in Refs.~\cite{arXiv-1208-3912,PRL-112-105301}\par
\begin{figure}[tb]
\includegraphics[width=250pt]{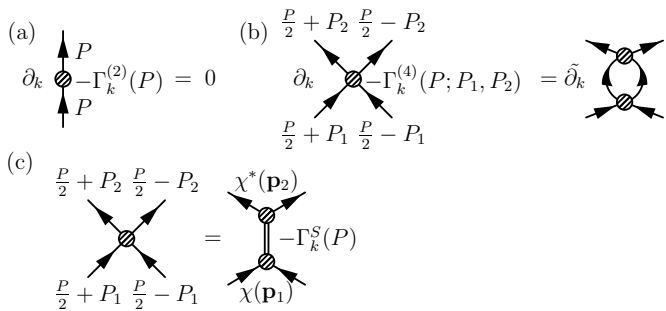}
 % 	\subfigure[]{\includegraphics[width=250pt]{1b2b.eps}\label{fig:1b2b}}
%	\subfigure[]{\includegraphics[width=230pt]{dec.eps}\label{fig:dec}}
%	\subfigure[Exact flow equation for three-body sector.]{\includegraphics[width=250pt]{feq3.eps}\label{fig:feq3}}
	\caption{Diagrammatic representations of the exact FRG equations for (a) the one-body sector and (b) the two-body sector. The derivative $\tilde{\partial_k}$ acts only on the internal propagator. (c) Decomposition of the 4-point 1PI vertex into the total-momentum and the relative-momentum parts.}\label{fig:1b2b}
\end{figure}
Based on this model, we perform an exact RG analysis based on FRG, which provides a non-perturbative RG scheme dealing with strongly correlated situations as Efimov physics. We start from the Wetterich equation \cite{PLB-301-90}:
\begin{eqnarray}
	\partial_k \Gamma_k[\psi,\psi^*]
		    =\frac{1}{2}\mathrm{Tr}\tilde{\partial_k}\mathrm{ln}
			\left(\frac{\delta^2\Gamma_k}{\delta\psi(q)\delta\psi^*(q)}+R_k({\mathbf q}) \right),\label{eq:wett}
\end{eqnarray}
where $\Gamma_k$ is the one-particle irreducible (1PI) effective action of the scale-dependent action $S_k=S+\int_Q R_k({\mathbf q})\psi^*(Q)\psi(Q)$ and reduces in the ultraviolet limit $k=\Lambda$ to the microscopic action $S$ and in the infrared limit $k=0$ to the usual effective action $\Gamma$, defined as the Legendre transform of the Schwinger functional. The symbol Tr implies the sum over momenta, Matsubara frequencies, and internal indices. The symbol $\tilde{\partial}_k$ acts only on the Litim's optimized regulator \cite{PRD-64-105007} $R_k({\mathbf q}):=(k^2-q^2)\Theta(k^2-q^2)$, where $\Theta$ is the unit-step function. %Note that the scale $k$ provides the cut-off scale of the RG, since $R_k$ serves as a $k$-dependent mass gap for a low-energy single-particle excitation.\par
To deal with the RG flow of the three-body coupling constant, we perform a vertex expansion \cite{PLB-334-355} of Eq.~(\ref{eq:wett}) with respect to the field variables to derive the RG equations for 1PI vertices:
\begin{eqnarray}
	\Gamma_k&=&\sum_{n=0}^{\infty}\frac{1}{(n!)^2}
	\int_{\substack{K_1,\cdots,K_n\\K'_1,\cdots,K'_n}}
	\Gamma_k^{(2n)}(K_1,\cdots,K_n;K'_n,\cdots,K'_1) \nonumber \\ 
	&&\times\delta(K_1+\cdots+K_n-K'_n-\cdots-K'_1) \nonumber \\
	&&\times\psi^*(K_1)\cdots \psi^*(K_n)\psi(K'_n)\cdots \psi(K'_1),\label{eq:vex}
\end{eqnarray}
where $\Gamma_k^{(2n)}$ is the $2n$th-order 1PI vertex, which represents the correlation of $n$ particles. Since we are interested only in the three-body physics, we have only to consider terms up to $n=3$. Indeed, the exact RG flow equations are closed up to $n=3$ since in the vacuum limit (i.e. the limits of diverging inverse temperature $\beta\rightarrow\infty$ and the vanishing number density of particles $n\rightarrow0$), the physics of four or more number of particles does not affect the three-body physics \cite{arXiv1301-6542}. In other words, in the vacuum limit, the diagrams containing particle-hole loops vanish because of the infinitely large chemical potential, which leads to decoupling of higher-order vertices from lower-order vertices, allowing an exact treatment of the RG equations. \par

%The exact RG equations therefore reduce to coupled equations of $\Gamma_k^{(2)}$, $\Gamma_k^{(4)}$, and $\Gamma_k^{(6)}$. 
We first consider one- and two-body sectors, which renormalize the three-body coupling constant. The exact RG equations in the vacuum limit for one- and two-body sectors are depicted in Figs.~\ref{fig:1b2b}(a) and \ref{fig:1b2b}(b), respectively. Noting the ultraviolet boundary condition $\Gamma_k=S$ $(k=\Lambda)$, we find that the one-body sector is given as
\begin{eqnarray}
	\Gamma_k^{(2)}(P)=ip^0+p^2-\mu,
\end{eqnarray}
which is consistent with the fact that the self-energy correction is absent in the particle vacuum. Because of the separate dependence on the relative momentum of the separable model, the two-body sector can be decomposed into the total-momentum and the relative-momentum parts as depicted in Fig.~\ref{fig:1b2b}(c), providing an analytical solution as follows:
\begin{widetext}
\begin{eqnarray}
&\displaystyle\Gamma_k^{(4)}(P;P_1,P_2)=\chi^*({\mathbf p}_2)\Gamma_k^S(P)\chi({\mathbf p}_1),\\
&\displaystyle\frac{1}{\Gamma_k^S(P)}=\frac{1}{16\pi a}-\frac{1}{2}\int\frac{{\mathrm d}^3l}{(2\pi)^3}
				\frac{ip^0+\frac{{\mathbf p}^2}{2}-2\mu
				+R_k(\frac{\mathbf p}{2}+{\mathbf l})+R_k(\frac{\mathbf p}{2}-{\mathbf l})}
				{\left[ip^0+\frac{{\mathbf p}^2}{2}+2{\mathbf l}^2-2\mu
				+R_k(\frac{\mathbf p}{2}+{\mathbf l})+R_k(\frac{\mathbf p}{2}-{\mathbf l})\right]2{\mathbf l}^2}
				|\chi({\mathbf l})|^2,
\end{eqnarray}
\end{widetext}
where $a$ is the $s$-wave scattering length.\par
\begin{figure}
\includegraphics[width=250pt]{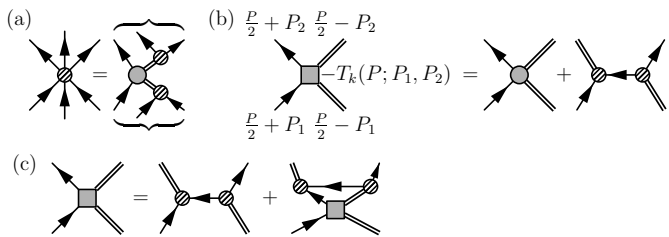}
\caption{(a) Decomposition of the six-point 1PI vertex, which describes the three-body scattering process. The double lines and the shaded vertices are the same as in Fig.~\ref{fig:1b2b}(c). The curly brackets mean symmetrization with respect to external lines. (b) Definition of the particle-dimer scattering amplitude $T_k$ represented as a square vertex. (c) Integral form of the exact FRG equation for the three-body sector. Integration of the RG equation with respect to the cut-off scale $k$ has been analytically performed.}\label{fig:feq3}
\end{figure}
The three-body sector can then be solved numerically based on these analytical expressions. Following a similar procedure as in Ref.~\cite{PTEP-2013-113A01}, we decompose the six-point 1PI vertex as described in Fig.~\ref{fig:feq3}. The exact RG flow equation for the three-body sector can then be analytically integrated with respect to $k$ and composed into a simple form as depicted in Fig.~\ref{fig:feq3}. We note that in the infrared limit $k=0$, this integrated RG equation reduces to the Skornyakov-Ter-Martirosyan equation \cite{JETP-4-648} for the separable model. Since we are only interested in the spatially isotropic $s$-wave component, which is relevant for Efimov physics, we make a projection onto $T_k(p,q):=\int\mathrm{d}\theta_{\mathbf{pq}}T_k(P^0_{\mathrm{onshell}}=3\mu;\mathbf{p},\mathbf{q})$, and define the dimensionless three-body coupling constant $g_3$ as a rescaled particle-dimer scattering amplitude as
\begin{eqnarray}
	g_3:=k^2T_k(p=0,q=0).
\end{eqnarray}
By solving the exact RG equation of the three-body coupling constant for the four different types of inter-particle interaction numerically, we obtain Fig.~\ref{fig:result}. We can see that the RG flows for the four different potentials show the interaction-dependent behavior at high energy; however, in the infrared regime, flows converge to the limit cycle. The onset point of the limit cycle is evaluated as $kr_{\mathrm{eff}}=0.49(4)$, in excellent agreement with the universal three-body parameter $\kappa^*r_{\mathrm{eff}}=0.49$ or $a_-/r_{\mathrm{eff}}=-0.43$. This observation suggests that the universality of the three-body parameter can be understood from the RG point of view as the universality of the onset of the limit cycle, which provides the first example relating the geometrical aspect of the limit cycle to the universal property of few-body physics. Our result also suggests that the three-body parameter can be regarded as the energy scale below which the discrete scale invariance of the system (including not only the Efimov states but also the periodic momentum dependence of the scattering observables) emerges.\par
\begin{figure}[tb]
  			\begin{center}
  			 	\includegraphics[width=100pt]{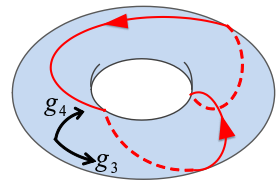}
				\caption{Schematic illustration of the limit cycle on the space of the three- and four-body coupling constants $g_3$ and $g_4$. A torus of $g_3-g_4$ space can be constructed by enclosing the space periodically. The limit cycle winds twice in the $g_4$ direction while it winds once in the $g_3$ direction. This reflects the fact that each Efimov state is associated with two four-body bound states \cite{PRA-70-052101,NP-5-417}. We suggest that such a nontrivial topology of the limit cycle may support the robustness of the number of bound states against a continuous change of the Hamiltonian.}\label{fig:lc4b}
  			\end{center}
\end{figure}

In this Letter, we have connected the missing link between two universalities of Efimov physics, namely the universal discrete scaling of the energy spectrum and the universal three-body parameter, by demonstrating that the renormalization-group limit cycle starts at the same point, regardless of short-range details.
% This point corresponds to the three-body parameter $kr_{\mathrm{eff}}=0.49$ consistent with experiments \cite{PRL-107-120401,PRL-108-145305,PRL-108-145305,PRL-111-053202} and numerical calculations \cite{PRL-108-263001,PRA-86-052516,arXiv-1208-3912,PRL-112-105301}.\par
%In summary, we clarified how the limit-cycle behavior is related to the universal three-body parameter; that is, the onset of the limit cycle in the renormalization-group flow is universally set by the finite-range correction. Our result is consistent with previous work Ref.~\cite{PRL-112-105301} which has concluded that the relevant length scale determining the three-body parameter is the effective range. The obtained value of  the three-body parameter $kr_{\mathrm{eff}}=0.49(4)$ is consistent with the experiments \cite{PRL-107-120401,PRL-108-145305,PRL-108-145305,PRL-111-053202} and numerical calculations \cite{PRL-108-263001,PRA-86-052516,arXiv-1208-3912,PRL-112-105301} within $15\%$ deviation. \par
An intriguing extension of the present work is to relate topological aspects of the limit cycle with universal properties of few-body physics. For example, when four identical bosons interact via a resonant interaction, two four-body bound states universally appear associated with one Efimov state exhibiting a universal scaling \cite{PRA-70-052101,NP-5-417}. We may relate this universal four-body bound states with a topological aspect of RG limit cycle. The previous RG analysis of four-body physics has shown that the four-body coupling constant $g_4$ forms a closed RG limit cycle which is solely induced by the limit cycle of the three-body coupling constant $g_3$ \cite{PRA-81-052709}. From this result we suggest that if we constitute a torus of the $g_3-g_4$ space by enclosing the space periodically, the closed limit cycle winds twice on the torus as schematically illustrated in Fig.~\ref{fig:lc4b}. Since the winding number of the limit cycle on the torus is topological, we may conclude that the number of four-body bound states is a topological winding number irrespective of the details of inter-particle interactions. This may afford a fundamental example that relates a topological property of a limit cycle to a universal property of few-body physics with scaling violation.\par
We acknowledge many fruitful discussions with Pascal Naidon, Endo Shimpei, and Yuya Tanizaki. This work was supported by KAKENHI Grant No. 26287088 from the Japan Society of the Promotion of Science, and a Grant-inAid for Scientific Research on Innovation Area ``Topological Quantum Phenomena" (KAKENHI Grant No. 22103005), and the Photon Frontier Network Program from MEXT in Japan. 

\end{document}